\documentclass[manuscript]{aastex}

\usepackage{amsmath}
\shorttitle{TRAPPIST-1: Planetary climates}
\shortauthors{Alberti et al.}

\begin{document}

\title{Comparative Climates of TRAPPIST-1 planetary system: results from a simple climate-vegetation model}

\author{Tommaso Alberti}
\affil{Dipartimento di Fisica, Universit\`a della Calabria, Ponte P. Bucci, Cubo 31C, 87036, Rende (CS), Italy}
\email{tommaso.alberti@unical.it, tommasoalberti89@gmail.com}

\author{Vincenzo Carbone}
\affil{Dipartimento di Fisica, Universit\`a della Calabria, Ponte P. Bucci, Cubo 31C, 87036, Rende (CS), Italy}

\author{Fabio Lepreti}
\affil{Dipartimento di Fisica, Universit\`a della Calabria, Ponte P. Bucci, Cubo 31C, 87036, Rende (CS), Italy}

\author{Antonio Vecchio}
\affil{LESIA - Observatoire de Paris, PSL Research University, 5 place Jules Janssen, 92190, Meudon, France}



\begin{abstract}
The recent discovery of the planetary system hosted by the ultracool dwarf star TRAPPIST-1 could open new perspectives into the investigation of planetary climates of 
Earth-sized exoplanets, their atmospheres and their possible habitability. 
In this paper, we use a simple climate-vegetation energy-balance model to study the climate of the seven TRAPPIST-1 planets and the climate dependence on the global albedo, 
on the fraction of vegetation that could cover their surfaces and on the different greenhouse conditions.
The model allows us to investigate whether liquid water could be maintained on the planetary surfaces (i.e., by defining a ``surface water zone'') in different 
planetary conditions, with or without the presence of greenhouse effect. 

It is shown that planet TRAPPIST-1d seems to be the most stable from an Earth-like perspective, since it resides in the surface water zone for a wide range of reasonable values of the model parameters. Moreover, according to the model outer planets (f, g and h) cannot host liquid water on their surfaces, even for Earth-like conditions, entering a snowball state. 
Although very simple, the model allows to extract the main features of the TRAPPIST-1 planetary climates.
\end{abstract}




\section{Introduction} \label{sec:intro}

The sharp acceleration of exoplanets discovery in the recent years \citep{NASA, Mayor95,Marcy96,Petigura13,Gillon16,Gillon17} and the presumed habitability of some of them 
\citep{Kasting93,Astrobiology,Spiegel08,Kopparapu13,Gillon17}, are changing our point of view on planetary science.

According to the usual definition \citep{Kasting93,Kopparapu13}, 
a planet resides in the so-called circumstellar habitable zone (HZ) if, being a terrestrial-mass planet with a CO$_\text{2}$-H$_\text{2}$O-N$_\text{2}$ atmosphere, it can sustain liquid water on its surface \citep{Kasting93,Kopparapu13}. The above requirements, coupled with the assumption of an Earth-like geology for the resulting greenhouse effect and carbon-silicate weathering cycle, 
imply that the surface temperature must be in the range $0 - 100$ $^\circ$C. 
Typically, apart from orbital features (e.g., eccentricity, period, transit time, inclination) and rough estimates of mass and radius, little information is directly known 
about exoplanets. 
For instance, the planetary surface temperature can be roughly estimated by using equilibrium conditions from energy-balance climate models depending on the 
distance of the planet from the hosting star and planetary outgoing energy. However, in such cases these estimates could be incorrect, since no information about the
planetary atmosphere is included in these models (as for Venus which has an estimated temperature of $\sim$ 300 K while the true surface temperature is about 737 K). 
Nevertheless, more complex energy-balance climate models, including the greenhouse effect and/or heat diffusion, provide insight into the climate on a planet
\citep[][and references therein]{Alberti15}.

Despite some recent comments on the metrics used to define the habitable zone in relation to public interest in scientific results \citep{comment1,comment2}, the recently 
discovered TRAPPIST-1 system \citep{Gillon16,Gillon17}, formed by seven temperate (with equilibrium temperatures $\lesssim$ 400 K) Earth-sized planets orbiting around a nearby 
ultracool dwarf, increased the attention to studying climate conditions of terrestrial exoplanets \citep{Bolmont17,Omalley17,Bourrier17,Wolf17}.
Since, as pointed out above, 
the estimates of equilibrium temperatures which do not take into account the greenhouse effect and the albedo feedback (so-called null Bond albedo hypothesis) cannot be sufficiently reliable, improved and advanced climate models, considering the planetary albedo and the atmospheric composition, are required \citep{Gillon17,Wolf17}. By using both a 1-D radiative-convective climate model and a more sophysticated 3-D model, \citet{Gillon17} found that inner planets, T-b, T-c, and T-d (in the following we indicate as T-x the x-th TRAPPIST-1 planet), show a runaway greenhouse scenario, while outer planets, T-e, T-f, and T-g, could host water oceans on their surfaces, assuming Earth-like atmosphere. Concerning the seventh planet T-h, due to the low stellar irradiance received, it cannot sustain surface liquid water oceans.
However, since only little is known about the planetary system, several approaches and hypothesis can be helpful in investigating both  planetary climates and atmospheric composition \citep{Devit16,Bolmont17,Omalley17,Wolf17}.

One of the drawbacks of the more detailed climate models is the necessary large pool of assumptions of atmospheric and surface conditions.
In this paper we investigate the possible climates of the TRAPPIST-1 planetary system by using a simple zero-dimensional energy-balance model \citep{Ghil15,Alberti15}, 
which allows the extraction of global information on the the climate evolution by using the actual knowledges about the planetary system. 
This model has the advantage of transparency through minimal assumptions, allowing a comparative sets of models to be studied. 
We study several situations, from completely rocky planets to Earth-like conditions, both neglecting and considering the greenhouse effect, to explore different
possible climates and make a comparative study of TRAPPIST-1 planetary system climates.

\section{The climate model} \label{sec:model}

The main features of the climate dynamics of Earth-like planets can be recovered through a zero-dimensional model based on two equations describing 
the time evolution of the global average temperature $T$ and of the fraction of land $A$ covered by vegetation:
\begin{eqnarray}
C_T \frac{dT}{dt} &=& \left[ 1-\alpha(T,A) \right] S(a,L_\star) - R(T) \label{eq:T} \; ,\\
\frac{dA}{dt} &=& A \left[ \beta(T) (1 - A) - \gamma \right] \; .
\label{eq:A}
\end{eqnarray}
Here $C_T$ is the planet heat capacity, $\alpha(T,A)$ is the planetary albedo, $S(a,L_\star) = L_\star/(4 \pi a^2)$ is the mean incoming radiation which depends on the star-planet 
distance $a$ (in au) and on the star luminosity $L_\star$, $R(T)$ is 
the outgoing energy from the planet, $\beta(T)$ and $\gamma$ are the vegetation growth and death rates, respectively \citep{Watson83,Ghil15,Alberti15}. The albedo of the planet depends on the fraction of land $p$, namely 
$\alpha(T,A) = (1-p) \alpha_o(T) + p [\alpha_v A + \alpha_g (1-A) ]$, where $\alpha_o$, $\alpha_v$ and $\alpha_g$ represent the albedos of ocean, vegetation and bare-ground, 
respectively. The albedo of the ocean is assumed to be linearly dependent on temperature as 
\begin{equation}
\alpha_o(T) = \alpha_{max} + \left(\alpha_{min} - \alpha_{max} \right) \left[ \frac{T-T_{low}}{T_{up} - T_{low}} \right]
\nonumber
\end{equation}
in a range of temperatures $T \in [T_{low}, T_{up}]$, resulting in $\alpha_o(T) = \alpha_{max}$ for an ocean completely covered by ice ($T \le T_{low}$) and $\alpha_o(T) = \alpha_{min}$ for an ice-free ocean ($T \geq T_{up}$).

The outgoing energy is described by a black-body radiation process, modulated by a grayness function, in order to take into account the greenhouse effect  
\begin{equation}
R(T) =  \left[ 1 - m \tanh \left( \frac{T}{T_0} \right)^6 \right] \sigma T^4 \; ,
\end{equation}
where $\sigma = 5.67 \times 10^{-8}$ W m$^{-2}$ K$^{-4}$ is the Stefan-Boltzmann constant, $m \in [0,1]$ is a grayness parameter ($m = 0.5-0.6$ for an Earth-like planet 
\citep{Sellers69,Alberti15}), and $T_0$ represents the mean global planetary temperature. The growth-rate $\beta(T)$ of vegetation is a quadratic function of temperature,
$\beta(T) = max \left[ 0; 1-k (T-T_{opt})^2\right]$ (being $k$ a parameter for the growth curve width and $T_{opt}$ an optimal temperature), while the death-rate $\gamma$ is 
assumed to be constant \citep{Watson83,Alberti15}. 

The free parameters $k$, $T_{opt}$, $\gamma$ and $\alpha_v$ are related to vegetation \citep[as in][]{Ghil15,Alberti15}. In the following, since we assume an Earth-like vegetation, these parameters are set to Earth's conditions.
A complete list of the used parameters and their corresponding values is shown in Table \ref{tab1}.

\begin{deluxetable}{cccccccccc}[ht]
\tablecaption{Values of the model parameters. \label{tab1}}
\tablecolumns{10}
\tablewidth{\textwidth}
\tablehead{
\colhead{Symbol} & & & & & & & \colhead{Value} & &  \colhead{Units}	\\
}
\startdata
$C_T$   	& & & & & & & 500 & & W yr K$^{-1}$ m$^{-2}$ 	\\
$\alpha_v$	& & & & & & & 0.1 & & 				\\	
$\alpha_g$	& & & & & & & 0.4 & &				\\
$\alpha_{max}$	& & & & & & & 0.85 & &				\\
$\alpha_{min}$	& & & & & & & 0.25 & &				\\
$T_{low}$	& & & & & & & 263 & & K				\\
$T_{up}$	& & & & & & & 300 & & K				\\
$T_{opt}$ 	& & & & & & & 283 & & K				\\
$k$		& & & & & & & 0.004 & & yr$^{-1}$ K$^{-2}$	\\
$\gamma$	& & & & & & & 0.1 & & yr$^{-1}$			\\
\enddata
\end{deluxetable}

As shown in previous studies \citep[][and references therein]{Sellers69,Watson83,Ghil15,Alberti15}, this set of parameters produces results in agreement with the observed Earth's surface temperature. In particular, the model also shows oscillatory solutions which can reproduce the observed sawtooth-like behavior of 
paleoclimate changes \citep[see][for more details]{Ghil15}.

By using the above set of parameters, we perform a parametric study of the solutions
as functions of the initial fraction $A_0$ of land covered by vegetation and of the bare-ground albedo $\alpha_g$. 
Moreover, we use different values of $p$ and $m$ in order to investigate the effect of land/ocean distribution and the role of the greenhouse effect on planetary climates.
We define a ``surface water zone (SWZ)'' as the circumstellar region where the 
planetary surface temperature ranges between $273$ K and $373$ K. It depends on the set $\{\theta\}$ of the variable parameters of the model and can be expressed as a a step-wise function
\begin{equation}
\text{SWZ}(\{\theta\}) = \left\{\begin{array}{cc} 
1 & \mbox{if  } 273   \leq T \mbox{K } \leq 373 \mbox{K}\; ,\\
0 & \mbox{otherwise} \; .\\
\end{array} \right.
\label{eq:Habit}
\end{equation}
Note that  $\text{SWZ}(\{\theta\})$, in the parameter space $\{\theta\}$, generally defines a range  where equilibrium temperatures calculated from the model are compatible with the presence of liquid water on planetary surface, independently from their atmospheric composition.

\section{TRAPPIST-1 planetary climates}

The possible climates of the TRAPPIST-1 planetary system are investigated by numerically solving Eq.s (\ref{eq:T})-(\ref{eq:A}) through a second order Runge-Kutta 
scheme for time integration and looking at the stationary equilibrium solutions \citep{Alberti15}.
The luminosity of the star is set to
\begin{equation}
S(a,L_\star) = \frac{0.0005}{a^2} S_\odot \; ,
\end{equation}
where, based on stellar properties of TRAPPIST-1 \citep{Gillon16,Gillon17}, we assumed that $L_\star = 0.0005 L_\odot$, $d_p = a * d_\odot$ (being $d_\odot = 1$ au the Sun-Earth distance), and $S_\odot = {L_\odot}/{4 \pi d_\odot^2} = 342.5$ W m$^{-2}$ is the mean solar radiation observed at the top of the Earth's atmosphere. 
The initial temperatures are set  equal to the equilibrium temperatures obtained by assuming a null Bond albedo \citep[see Table 1 in][]{Gillon17} 
and the scale parameter $a$ is chosen as the mean distance of each T-x planet to the TRAPPIST-1 star \citep{Gillon17}. 

\begin{figure*}[ht]
\plotone{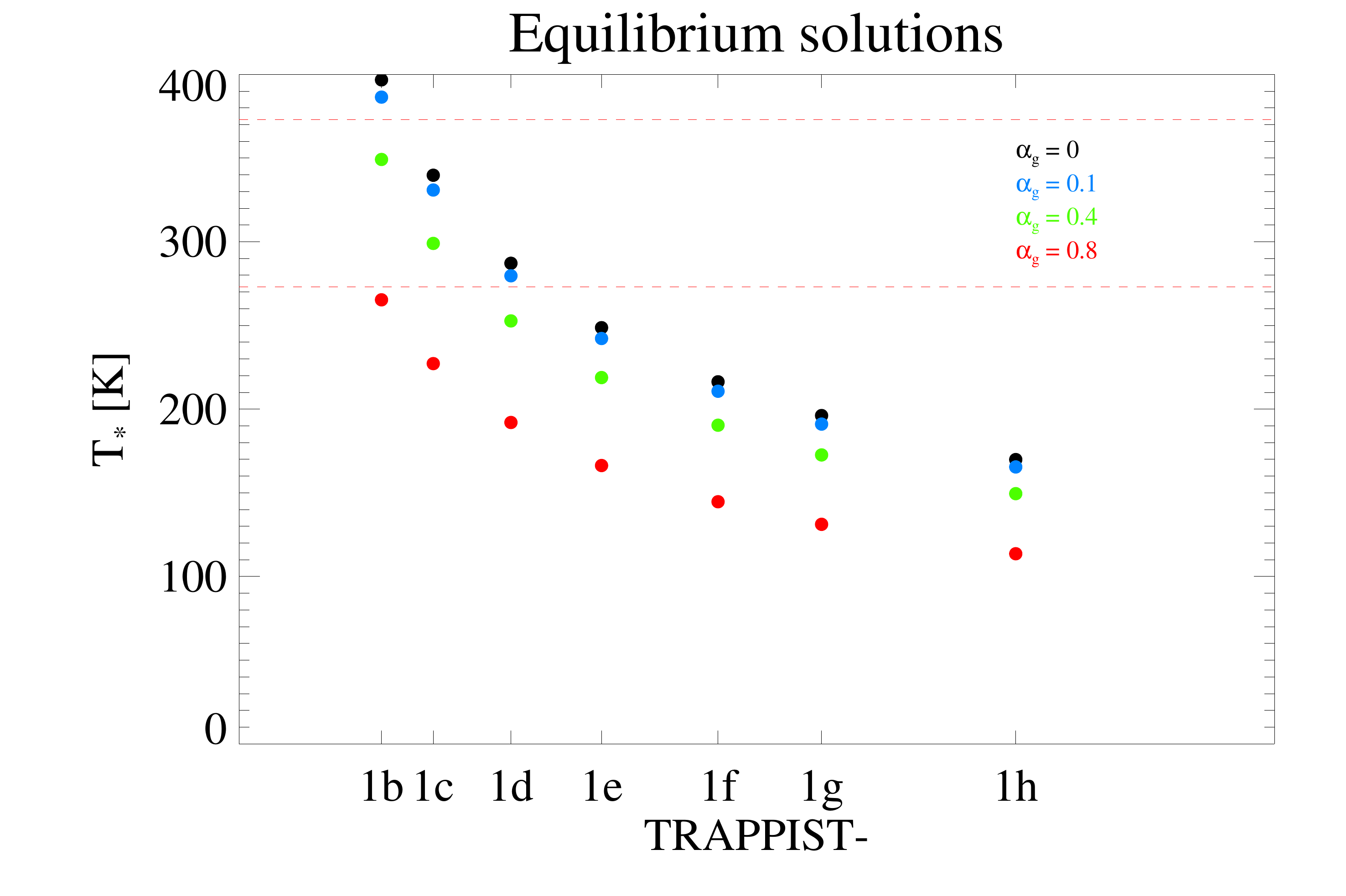}
\caption{Equilibrium solutions for temperatures as functions of the star-planet distance for four values of $\alpha_g$. Values refer to rocky planets without vegetation and 
greenhouse effect ($p=1$, $A=0$, $m=0$). The red dashed lines represent the temperature range for which planets are in the SWZ.}
\label{fig:fig1}
\end{figure*}

First of all we consider the case of rocky planets ($p=1$) with no vegetation ($A=0$) and no greenhouse effect ($m = 0$).
In Figure \ref{fig:fig1} we show the stationary solutions for the temperature of the planets, obtained from Eq.s (\ref{eq:T})-(\ref{eq:A}), as functions of the star-planet distance, 
for different values of the bare-ground albedo $\alpha_g$. 

When $\alpha_g$ is set to zero 
(i.e., for black dots in Figure \ref{fig:fig1}), solutions reported in Table 1 of \citet{Gillon17} are obtained, since they are equilibrium solutions of an energy-balance 
climate model with null Bond albedo hypothesis. Moreover, in this case only
T-c and T-d reside in the surface water zone (SWZ). 
However, as the bare-ground albedo $\alpha_g$ is changed, different conditions can be observed, that is, as $\alpha_g$ increases some T-x planets can enter in or exit from the 
surface water zone. For example T-b could host surface liquid water on its surface only for a range of $\alpha_g$ close to $\alpha_g \simeq 0.5$. 
This suggests that in the simple case when planets are mainly rocky and without atmospheres, their residence in the surface water zone is dependent on their surface albedo.
This consequently implies that the vegetation coverage is a main feedback acting as a thermal regulator for planetary temperature.

For the above reasons, in the following we investigate the  climate properties of the planetary system when planetary conditions, related to different surface 
vegetation coverage and bare-ground albedo, are  considered. 
We will show, in detail, T-x temperatures for three different situations: i) rocky planets without oceans or ice ($p=1$) and without greenhouse effect ($m=0$); ii) Earth-like land distribution ($p=0.3$) without greenhouse effect 
($m = 0$); iii) Earth-like land distribution ($p = 0.3$) with a greenhouse effect similar to that observed on Earth ($m = 0.6$).
This gradual approach is useful to investigate planetary climates starting from different conditions, in order to make a comparative study on the possible climates of TRAPPIST-1 planets by considering several possible situations.

\begin{figure*}[ht]
\plotone{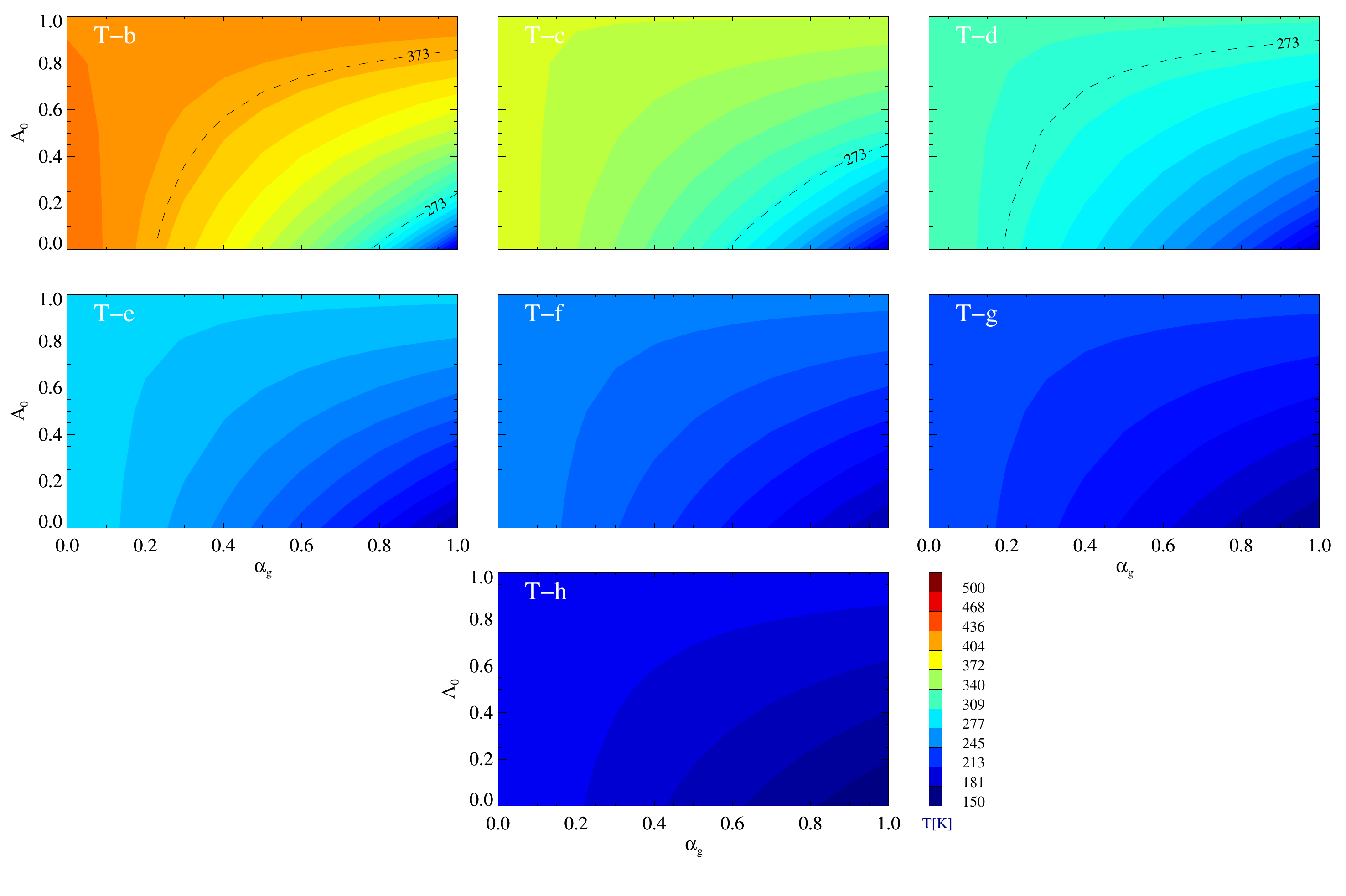}
\caption{Equilibrium solutions for temperatures in the plane $(\alpha_g,A_0)$, for rocky planets without vegetation and greenhouse effect ($p=1$, $m=0$). 
The surface water zone, when present, is shown by dashed lines}.
\label{fig:fig2}
\end{figure*}

The stationary solutions for temperatures in the plane $(\alpha_g,A_0)$ are shown in Figure \ref{fig:fig2} for the case of rocky planets ($p=1$).
As said before, only the first three planets can be in the surface water zone when no atmosphere is considered ($m=0$). Moving from low to high values of $A_0$, 
planetary surface temperature changes accordingly, increasing as $A_0$ increases. 
More specifically, for $A_0 \ge 0.5$ T-c displays only a weak temperature  variation with $\alpha_g$. This indicates that vegetation acts as
a feedback to maintain conditions for which $\text{SWZ}(\{\theta\})=1$. A similar behavior is recovered for T-d but for higher values of $A_0$ ($A_0 \gtrsim 0.8$), suggesting that this exoplanet should 
be almost completely covered by vegetation to reside in the surface water zone. Conversely, due to its lower distance from the star, T-b shows an opposite behavior, 
entering the surface water zone for lower values of $A_0$, namely $A_0 \lesssim 0.5$. For planets T-e, T-f, T-g and T-h, although the equilibrium temperature increases with $A_0$, the stellar irradiance is not enough to have global temperatures
compatible with a surface water zone. We remark that the presence of liquid water in a planet without atmosphere does not make intuitive sense. The atmospheric envelope is indeed a fundamental component of a climate system to develop and maintain conditions for life on a planet.
Therefore the case just discussed is presented mainly for a comparison of surface temperature conditions with the more Earth-like cases which are reported below.

\begin{figure*}[ht]
\plotone{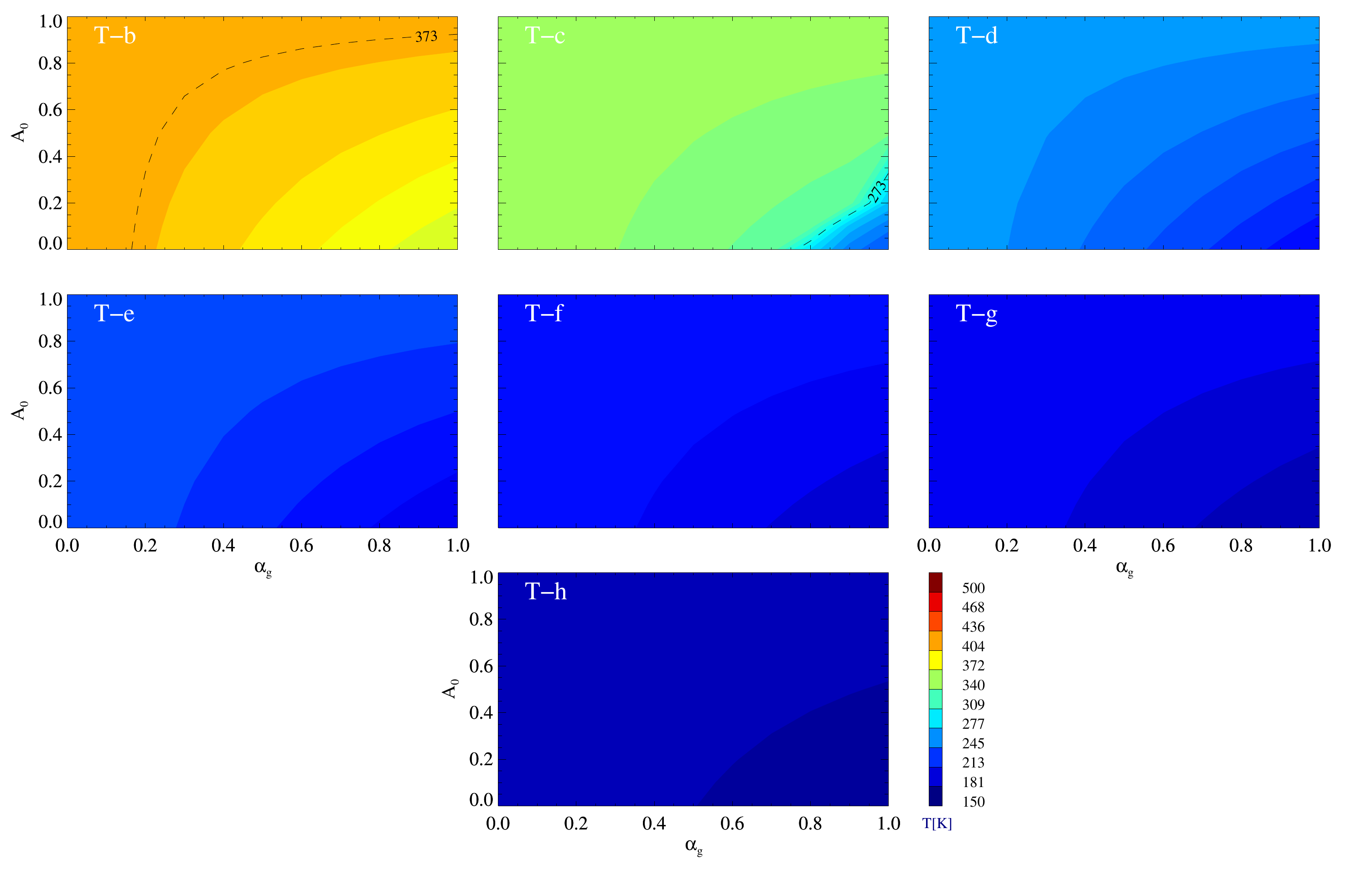}
\caption{Equilibrium solutions for temperatures in the plane $(\alpha_g,A_0)$, for planets with a fraction of land and oceans similar to the Earth ($p=0.3$). Greenhouse effect is 
not included ($m=0$). The surface water zone, when present, is shown through dashed lines.}
\label{fig:fig3}
\end{figure*}

We turn now to a situation where planets have an Earth-like land distribution. This is done by setting $p=0.3$, corresponding to a planet covered by land, ocean and ice. We also assume 
that the greenhouse effect is negligible ($m=0$). Figure \ref{fig:fig3} shows the stationary solutions for temperatures when $p=0.3$ and $m=0$.
The main difference with the previous case concerns planet T-d. Indeed, when an Earth-like land distribution is considered, the global surface temperature is always lower than $273$ K, that is, T-d is not in the surface water zone. 
For both T-b and T-c there are wide ranges of the parameters for which $\text{SWZ}(\{\theta\})=1$, while the other planets cannot host surface liquid water on 
their surfaces. The observed changes in equilibrium temperatures suggest that oceans play a primary role in setting the thermal equilibrium conditions for planetary surface 
temperature, at least when the greenhouse effect is not considered. 

\begin{figure*}[ht]
\plotone{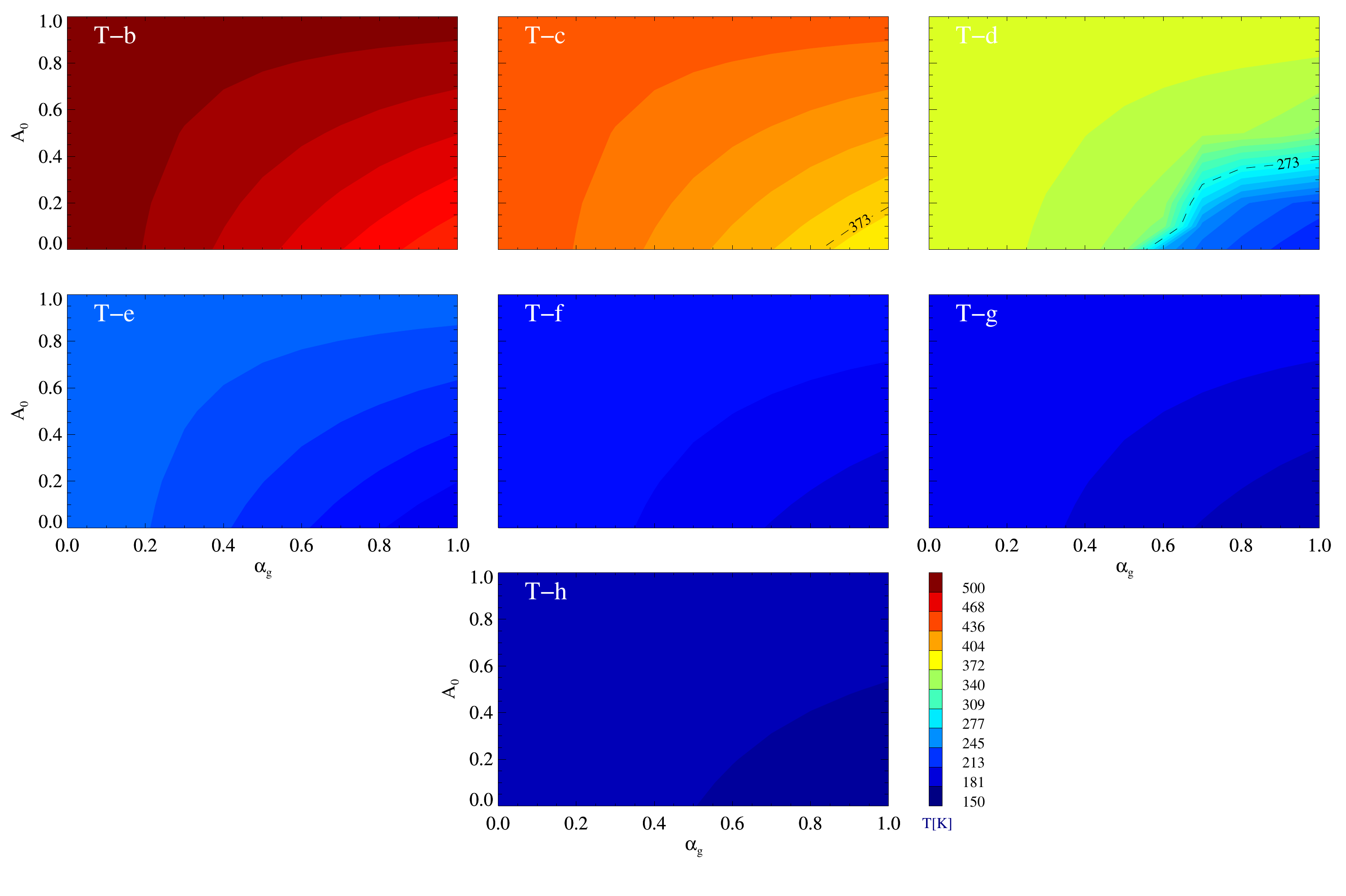}
\caption{Equilibrium solutions for temperatures in the plane $(\alpha_g,A_0)$, for planets with an Earth-like fraction of land and oceans and greenhouse effect ($p=0.3$, $m=0.6$). 
The surface water zone, when present, is shown through dashed lines.}
\label{fig:fig4}
\end{figure*}

Let us now consider the same  land distribution ($p=0.3$), but with an Earth-like greenhouse ($m=0.6$). The stationary solutions for temperatures are shown in Figure 
\ref{fig:fig4}. 
The planetary climates change with respect to the previous case, since temperatures increase. In particular, T-d again resides in the surface water zone for a wide range of $A_0$ and 
$\alpha_g$, although for $A_0 < 0.5$ the planet leaves that zone when $\alpha_g > 0.5$. 
As for the previous cases, even using the Earth's value of $m$, outer exoplanets cannot reach global surface temperatures in the range [273~K,373~K]. This implies that these planets 
need a different greenhouse effect with respect to the Earth, and consequently a different atmospheric composition, to enter the surface water zone. 
These results are quite different from those by \citet{Gillon17} who showed that outer planets, with Earth-like atmospheres, could host water oceans on their surfaces.
This could be related to the main difference between our model and that used in \citet{Gillon17}. While in our model the greenhouse effect is included by using a parametric approach \citep{Sellers69,Alberti15}, \citet{Gillon17} utilized
a 1-D radiative-convective cloud-free model in which greenhouse effect is taken into account by considering the contribution of several types of greenhouse gases (e.g., CO$_\text{2}$, H$_\text{2}$O, N$_\text{2}$) with different partial pressures \citep{Wordsworth10}.
However, our results are quite in agreement with 
that reported by \citet{Wolf17}, who showed, by using a 3-D climate model, that outer planets (i.e., T-f, T-g, and T-h) are not warmed enough, falling beyond the habitable zone and 
entering a snowball state.
On the other hand, inner planets T-b and T-c show higher surface temperatures, so that to
reside in the surface water zone their greenhouse effect should be similar or
less efficient than on the Earth. In particular, T-b cannot be in the surface water zone for Earth-like greenhouse effect, while, when low values of $A_0$ and high value of $\alpha_g$ are considered, T-c is in the surface water zone for a narrow range of the model parameters.
These results on inner planets are also in 
agreement with \citet{Wolf17} who reported that the inner three planets (T-b, T-c, and T-d) could reside in the traditional liquid water habitable zone but only with runaway greenhouse 
conditions.

The surface water zone changes significantly with the planetary atmospheric composition, such that, for instance, the first three planets can reside in the surface water zone 
for a wide range of the parameters, or, in some cases, even none of them can be there. In particular, it is interesting to see whether T-e, which is at the center of the system, 
could enter the  surface water zone, since previous studies by \citet{Wolf17} have suggested this planet has the best chance to have water oceans on its surface. 
For this reason, we investigate changes in the surface water zone keeping fixed $\alpha_g = 0.4$ and  varying both $p$ and $m$.

\begin{figure*}[ht]
\plotone{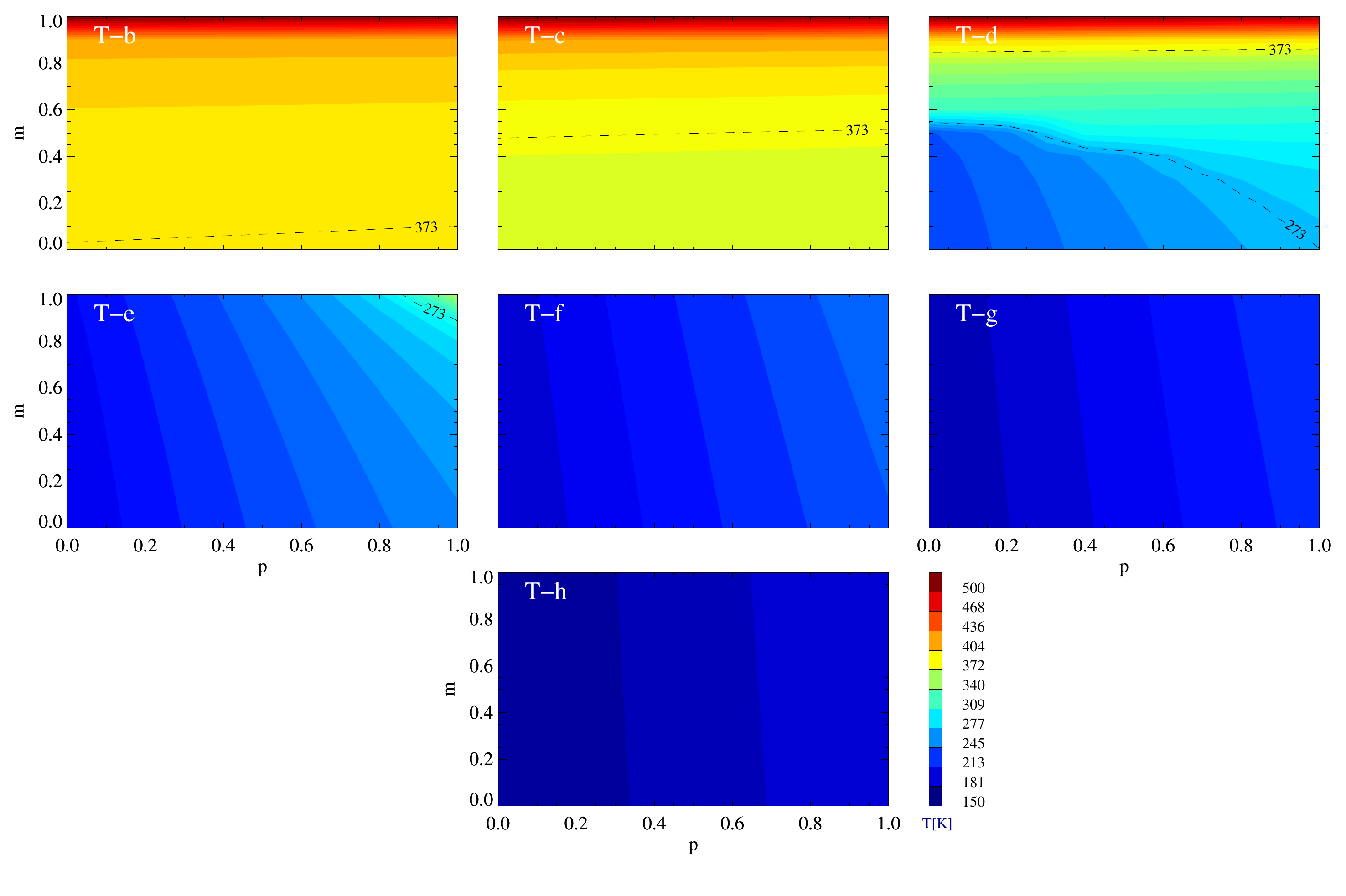}
\caption{Equilibrium solutions for temperatures in the plane $(p,m)$ for Earth-like bare-ground conditions ($\alpha_g=0.4$). 
The surface water zone, when present, is shown by dashed lines.}
\label{fig:fig5}
\end{figure*}

In Figure \ref{fig:fig5} we show the stationary solutions for temperatures in the plane $(p,m)$. As expected, the planetary surface temperature strongly depends on the greenhouse effect conditions.
In particular, T-b has a very narrow range of surface liquid water for very low values of $m$, while planets T-c and T-d have a wide range of parameters for which $\text{SWZ}(\{\theta\})=1$. 
Interestingly, for high values of $p$ and $m$, T-e enters the surface water zone. In an Earth-like situation this planet cannot have surface liquid water for lower values of $m$, 
suggesting that its atmosphere must have higher levels of greenhouse gases, such as CO$_\text{2}$ or N$_\text{2}$, with respect to those observed on Earth. 
This result is quite different from that obtained by \citet{Wolf17} according to which T-e has the best chance to be a habitable ocean-covered planet.
This discrepancy could be related to the fact that our model considers the greenhouse effect in a parametric way, while the 3-D model used in \citet{Wolf17} directly uses the contribution of several types of 
greenhouse gases, as N$_\text{2}$ and CO$_\text{2}$ \citep[similarly to][]{Wordsworth10}. 
Indeed, a zero-dimensional climate model can determine the effective planetary emissivity of long wave radiation emitted to space, while a radiative-convective model considers 
different processes of energy transport, from radiative transfer through atmospheric layers to heat transport by convection. This allows to directly investigate the effects of 
varying greenhouse gas concentrations on thermal energy balance.
However, although the results are different, the number of unknowns is such that it is not possible to know which climate model is more likely.
Finally, outer planets (T-f, T-g and T-h) seem to be not in the surface water zone, entering a snowball state \citep{Wolf17}, even if the greenhouse effect increases to
higher levels than those observed on Earth.

\section{Conclusions}

In this paper we investigated the climate of the TRAPPIST-1 planetary system by using a  zero-th order energy-balance model which allows us to outline the main features 
of the different planets. We found that the surface water zone, defined as the circumstellar region where a planet can host liquid water on its surface, is strongly dependent on the different parameters of 
the model and, in particular, on the initial fraction of vegetation coverage, the bare-ground albedo and the presence of oceans. More specifically, the ``inner'' three 
planets T-b, T-c, and T-d seem to be located in the surface water zone for several values of the parameters, as described before, while planet T-e, at variance to what have been 
reported in \citet{Gillon17}, can present water oceans only for greenhouse effect conditions different from the Earth. 
The climate of planet T-d seems to be the most stable from an Earth-like perspective, because this planet resides in the range of $\text{SWZ}(\{\theta\})=1$ for a wide interval of reasonable values 
of the different parameters. This result is not in agreement with that reported by \citet{Wolf17} for which the best candidate for a habitable ocean-covered surface is the 
planet T-e. This difference could be related to the different models employed, as in our energy-balance model a parametric description of the greenhouse effect is used, 
while in the 3D climate model by  \citet{Wolf17} the contribution of several types of greenhouse gases is taken into account. 
However, since the number of unknowns makes difficult to choose one model with respect to another, different approaches, based either on simple or more complex 
climate models, can be useful. 
In this framework, our model has the advantage of transparency through minimal assumptions, allowing a comparative sets of cases to be studied. 

Here we showed that the TRAPPIST-1 system can have different climates and that equilibrium temperatures depend on the global albedo, that is,  on the mean physical conditions of the planetary surface. However, this parameter is strongly variable, since the 
vegetation could cover only a fraction of the surface, as for example the case of Earth. 
Moreover, also the greenhouse effect needs to be properly considered since it is one of the main feedback in regulating thermal energy balance.
Investigating these features require more sophisticated models, extended to space variables, at least with a description of the atmospheric heat diffusion. The model is actually 
under investigation and results will be reported in a forthcoming paper.

\acknowledgments
We acknowledge S. Savaglio for useful discussions and for her interest in our work. 
We thank the anonymous reviewer for fruitful and helpful suggestions.

\end{document}